\documentclass[11pt]{article}
\usepackage[english]{babel}
\usepackage[latin1]{inputenc}
\usepackage{url} 
\usepackage{framed}
\usepackage{graphicx}
\graphicspath{{images/}} 
\usepackage{amsmath}
\usepackage{array}
\usepackage{fullpage}
\usepackage{setspace}
\begin{document}

\title{
  Data, Science and Society\\ \medskip
\small (Notes for a talk at LEARN Conference, May 2017)\footnote{
Data, Science, Society. Talk at LEARN Final Conference,
 Senate House, University of London,  London, May 5th, 2017.} \\
Version 1.0, May 4th, 2017
}
\author{Claudio Gutierrez \footnote{ 
Computer Science Department and Center for Web Semantics,
 Universidad de Chile, Chile.}
}
\date{}
\maketitle

\begin{quote}
 {\em The foundations of experience (since we absolutely must get down to
this) have been non-existent or very weak; nor has a collection or
store of particulars yet been sought or made, able or in any way
adequate, either in number, kind or certainty, to inform the
intellect. 
[...] Natural history contains nothing that has been researched in the
proper ways, nothing verified, nothing counted, nothing weighed,
nothing measured.}

Francis Bacon, 1620.\footnote{
F. Bacon, The New Organon. Edit. L. Jardine,
M. Silverthorne. Cambridge Univ. Press, 2000.
Aphorism XCVIII, p. 80.}
\end{quote}

The word ``data'' comes from the plural of the Latin word {\em datum},
meaning something given.\footnote{
Despite that data is hlghly countable, in what follows, I will consider
 ``data'' as non-countable noun, like
water, love, information, oil. Hence I will treat it as singular.
} The first uses of the word in a scientific
context can be traced back to the mid 17th century, where the
quote by Bacon suggests it arose as a means to try to conceptualize
scientific research. In modern-day usage, the word refers to
collections of measurements and factual information that form a basis for
research, reasoning, supporting evidence, etc.\footnote{
D. Rosenberg. Data Before the Fact. American Historical Association, 2012.
}
With the advent of computers in the mid 20th century,
a new meaning was added to this traditional scientific sense.
It indicates the basic abstract entities that the new machines deal
 with (``transmittable and storable computer information'', 1946).
 Both, the scientific and computing senses remained  confined to rather
technical communities until recently. For example,  the word ``data'' did not capture enough attention from
Raymond Williams to merit its inclusion in {\em Keywords}  in 1983.\footnote{
R. Williams. Keywords, 1983.
  }

   The  popularization of the  word ``data'' in the news headlines,
 in   magazine covers  and in any report  
 seeking to be considered scientifically grounded,  is quite recent. 
 Two metaphors are responsible for this miracle: first,  the notion of
 ``data deluge'' and then the noun ``big data''.


 The notion of data deluge is powerful: an overflow of society and humans with data. 
 Despite its strong call to our daily experience,
 the notion is highly misleading. 
First, it suggests something produced by an other, traditionally
 by a punishing  God, or by natural powers external to our command.
Second, it makes current levels of data a catastrophe,  surrounding
data with a negative connotation.
 In summary, it presents data as something that we can only react to,
or even  defend ourselves from.

 The notion of big data is less deceiving. It avoids the explicit negative
 connotation and highlights a main feature of the phenomenon: the size. 
People in computing coined the term in the early nineties\footnote{
See
\url{https://bits.blogs.nytimes.com/2013/02/01/the-origins-of-big-data-an-etymological-detective-story/}
},
but the hype in bussiness (that popularized it) is more recent.
In science and reaserch the term  began to be widely adopted only in
the 21st century.\footnote{
Editorial of Nature special issue about Big data (2008)
``Researchers need to adapt their institutions and practices in response to torrents of new data''.
}
My concern with this notion is that it still represents the phenomenon
as an external and unapproachable object.
In fact, many people speak (and think) of ``big data'' as an
 obscure and phantasmal object, far from the control of mankind,
 haunting modern society.
It is like a living nebula growing around us 
waiting to be tamed and ready to run over us if not taken seriously.

\section{Torrents of Data}

Let us retain the essential fact:
there is an enormous amount of data around.
At this point one can ask why this hype over data {\em now}.
Historically, there have been several  tsunamis overflowing the semantic
and symbolic capacities of the human.
The adoption of writing and of media to
preserve it without doubt shook and transformed contemporary ways of
approaching information. Later the printing press must have produced a similar
upheaval among the literate population. More recently,  magazines, newspapers
and pervasive technology of printing, plus the radio and television
media overhelmed people with information. 
 In the 1930's, the Spanish thinker
Jos\'e Ortega y Gasset, in a speech to the International Congress
of Libraries and Bibliography, spoke of the 
``raging book'' and expressed his fears as follows:
\begin{quote}
``There are already too many books. Even when we drastically reduce the
number of subjects to which man must direct his attention, the
quantity of books that he must  absorb is so enormous that it exceeds 
the limits of his time and his capacity of assimilation.
[...]
The culture which has liberated man from the primitive forest now
thrusts him anew into the midst of a forest of books no less
inextricable  and stifling.'' And concluded:
``Here then is the drama: the book is indispensable at this stage in
history, but the book is in danger because it has become a danger 
for man.''\footnote{
Jos\'e Ortega y Gasset. Misi\'on del Bibliotecario. 1935.
The Mission of the Librarian. Translated by J. Lewis and R. Carpenter.
The Antioch Review, Vol. 21, No. 2, 1961, pp. 133-154.
}
\end{quote}

The book --according to Ortega-- became a danger for man.
We hear similar remarks
about data today. What is going on beneath the surface? The problem
then and today can be stated, paraphrasing a well known text about social
change, as follows:
{\em At a certain stage of development, the material forces of society
began producing more symbolic material than the one existing
social relations can digest. From forms of development of the
culture these relations turn into their fetters.
Then begins an era of information upheaval.}
The problem we face today is that capture, production and
digestion of data surpasses by far the social and human capabilities
to manage and process it.

 Let us develop this idea. 
The data that worries us are not billions of units of understandable
data (what scared Ortega), but the not understandability of the unit itself.
Let us explain. There are no unreadable books. 
Each of them was designed (written) to the be read by a human
(even {\em In Search of the Lost Time}).
The obstacle Ortega was pointing to was the  almost infinite number of
books.
The main problem was the quantity (and his solution functioned
accordingly:  to limit the production of books).
On the other hand, the problem with data is that the unit itself
(a dataset) is not intelligible by a human, and worst, there are almost
infinitely many datasets. The problem is both quality and quantity.


A metaphor from physics could help at this stage. 
Traditional mechanics  is a {\em human-scale} discipline in the sense
that allows the direct participation of people. A bicycle is an
artifact  we can understand, repair, transform almost entirely
by ourselves. 
Chemistry first and then atomic physics  crossed the barrier of the
human manageable objects, a barrier beyond which the built-in
senses  of the human do not help anymore.
 Today the advances in new technological
 media (computer power and memory, networks,
 sensors, communication, etc.) 
 has dramatically increased the capacity of capturing data 
(sensors, telescopes, Web, etc.); of producing data (computers,
games, media, LHC, etc.); of storing data (memory, storage media, cloud, etc);
of analyzing data (statistical techniques, neural networks,
(deep) learning, etc.).
In one sentence: the limit we are surpassing today is that
of the human capabilities to understand and manipulate these
vast new world of symbolic objects.\footnote{
Hans Moravec. When will computer hardware
match the human brain? Journal of Evolution and Technology. 1998. Vol. 1.
\url{http://www.jetpress.org/volume1/moravec.htm}
}

%
%
%
%
%
%




 Despite Ortega's complaints, until very recently we
humans could approach all symbolic objects around us: texts, photographs,
music, movies. But today this symbolic world is growing so fast that
escapes our ``natural'' human and societal capacities to handle it, and 
thus we feel that an obscure and  daunting, fundamentally unintelligible, (parallel) world is
growing in front of our eyes.
 Let us clarify. It is not that this symbolic world did not
exist before. It existed and was vast, but essentially volatile.
The essential novelty 
 is that it is being increasingly materialized in the form of
 (digital) data and that ICT technologies have made us become
aware of its vastness. (See some numbers in table 1.)
 My hypothesis is that these objective and subjective phenomena have
 made obsolete the conceptual models used to deal with the symbolic world.
 Among the main challenges is the notion of scale.\footnote{
Clark C. Gibson, Elinor Ostrom, T.K. Ahn.
The concept of scale and the human dimensions of global change: a
survey.
{\em Ecological Economics} 32 (2000) 217-239.
}

\begin{table}[!t]
  \center
 \[
   \text{human scale}  
  \begin{cases}  &  
\begin{tabular}{lcl} 
   Byte B    &     $\sim 10^0$        & a character  \\
    Kilo KB   &  $\sim10^3$ & written  text \\
   Mega MB  &  $\sim 10^6$           & image, music  \\
   Giga  GB  &    $\sim 10^9$         & movies \\ 
\end{tabular} \\
  \end{cases}
  \]
  \[ 
 \hspace{1.2cm}\text{beyond human}
  \begin{cases}  & 
  \begin{tabular}{lcl}
  Tera TB   &     $\sim 10^{12}$         &  U.S. Congress Library \\       
  Peta PB   &     $\sim 10^{15}$         &   Large data center     \\
  Exa  EB  &       $\sim 10^{18}$        &    All words ever spoken  \\
  Zetta ZB &       $\sim 10^{21}$        &    Amount of global data 
  \end{tabular}
  \end{cases}
  \]
\caption{Data sizes and human-scale}
\end{table}

\section{The Notion of Data}

 The audience --mostly of librarians, those rare intellectuals whose 
life's objective is to contribute to other people's growth--
is  familiar with the subtle relations existing  among
the notions of knowledge and information.
Let us assume for now the widely accepted premise  that data is in some
sense the starting point,  constitutes a basic building block, of
information  and knowledge.\footnote{
Ch. Zins. Conceptual approaches for defining data, information, and knowledge. Journal of the American Society for Information Science and
Technology 58, 2007. pp. 479-493
} 
On these lines, let  us present some facets and assumptions relevant
for our  understanding of this concept.

\medskip

 1.  At the most basic and abstract level, data is a distinction, {\em i.e.},
a sign of a lack of uniformity in the world out there.
As Luciano Floridi states: 
``As ``fractures in the fabric of
being''  they can only be posited as an external anchor of our
information, [...] are never accessed or elaborated
independently of a level of abstraction.''\footnote{
L. Floridi, Semantic Conceptions of Information.
SEP,  version Wed Jan 7, 2015.
\url{https://plato.stanford.edu/entries/information-semantic/}
}
Caroline Haythornthwaite points to the same from another point of view: 
``Datum is [the] smallest collectable unit associated with a phenomenon.
Normally, data occur in collections that are collected in
order to monitor a process, assess a situation, and/or otherwise gain
a referent on a phenomenon.'' 
\footnote{
C. Haythornthwaite. In Ch. Zinns, {\em op. cit.}, p. 483.
}
In summary, data is the most basic layer in the symbolic world. 
Data has no meaning by itself, but is the source of meaning.

\medskip

2.  By data, we will mean materialized (digitally recorded) data,
   that is, data once it has been frozen into material (digital)
   symbols.
 In this regard data,
in the sense we are
treating it in these notes,  is part of the ``objective'' world.
Data is thus a material collection of symbols. This is the spirit of the
following entries in the definition of data: 
 ``information in numerical form that can be digitally transmitted or
 processed'' (Merriam-Webster) and 
``The quantities, characters, or symbols on which operations are
performed by a computer, which may be stored and transmitted in the
form of electrical signals and recorded on magnetic, optical, or
mechanical recording media.'' (Oxford).
%
%
In summary, despite its ontological ambiguity between the
material and the intangible,  data is material. 

\medskip

3.  The distinctions that define data assume an implicit  context. 
 This network of meanings is not stated explicitly, that is,
not specified in the data itself. This allows 
manifold interpretations of the same data from different
points of view, to further explore new dimensions, etc.
A good example is a photograph. With high probability, the
photographer took it with some agenda in mind. But future
generations could use it to ``view'' dimensions that were
not present in the original focus of the original photographer.
Some contexts are usually explicitly included as metatada,
that is, additional data that give information or signal relations in the bulk data.\footnote{
  It is important to distinguish the intrinsic metadata, what in the
  database field is called a {\em schema} (describing existing types, relationships
  among fields, etc. in a dataset), from metadata describing whole datasets.
 Example of the latter are the 15 classical properties of the
  Dublin Core Metadata Element Set (version 1.1):
contributor, coverage, creator, date, description, format, identifier,
language, publisher, relation, rights, source, subject, title, and type.
}
   In summary, data has meaning, though not always explicit meaning.
Although collected/constructed with some objective in mind, 
it allows diverse interpretations and can support manifold thesis.


\medskip

%


 Data is the starting point for our discussion.
Our task is not to clarify the ontological status of data, but to understand its properties, its ``mode of combination'', and hopefully to get a conceptual model for it.
  That is, for us, as data guardians, curators, facilitators,
data is just something given, as the original Latin meaning.
Our concern at this stage is not the
 possible semantics that could be distilled from the data, 
 but the data as ``material''  element. 
 Using the counterpoint between worlds of bits and atoms
  popularized by Nicholas Negroponte in {\em Being Digital},\footnote{
N. Negroponte. Being Digital. Vintage Books, New York, 1996. 
}   
   we work in the {\em world of bits}, a world as material as the one of
   atoms, {\em but} with strongly different social significance as we will see.
  
  Taking advantage of the bit-atom opposition,
 another metaphor could help shedding light
on the relations between these two worlds:
\[
 \frac{\text{Data}}{\text{Virtual World} } = 
\frac{\text{Atoms}}{\text{Material world}} 
\]
Pushing forward the association of ideas,
 the science of data would be the chemistry
of the virtual world. The sciences of information and knowledge work
with this material, but at a different  level of grouping and abstraction.

\section{Research and scientific Data}


 The notion of research data, under  terms like  experience, facts,
 observation, evidence,  etc. has a long history. 
 ``Observation'' in its scientific sense is mentioned by Aristotle;
Bacon advocated its relevance for research; and the awareness
of  the subleties of its connection to knowledge date from 
the beginning of the 20th century.\footnote{
James Bogen, Theory and Observation in Science.
SEP, version Mar 28, 2017.
\url{https://plato.stanford.edu/entries/science-theory-observation/}
 }
    Nevertheless, is it only at the turn of the 21st century that data
    began to be thought of as a driver of science.  
Turing award recipient Jim Gray wrote in 2007:


\begin{quote}
``Originally, there was just experimental science, and then there was
theoretical science, with Kepler's Laws, Newton's Laws of Motion,
Maxwell's equations, and so on. Then, for many problems, the
theoretical models grew too complicated to solve analytically, and
people had to start simulating. These simulations have carried us
through much of the last half of the last millennium. At this point,
these simulations are generating a whole lot of data, along with a
huge increase in data from the experimental sciences. [...]
The world of science has changed, and there is no question about this.''
\footnote{
Jim Gray on eScience: A Transformed Scientific Method.
(Based on the transcript of a talk given by Jim Gray
to the NRC-CSTB1 in Mountain View, CA, on January 11, 2007.)
In: 
 T. Hey, S. Tansley, K. Tolle.
  The Fourth Paradigm. Data-Intensive Scientific Discovery.
  Microsoft Research, 2009.
}
\end{quote}

 This change driven by the ``material forces of society'' is producing
 social changes, in particular giving a prominent value to scientific
 data. 
The argument works as follows.

 Since the industrial revolution there was
awareness of the expanding role of (scientific) knowledge in the 
economy, but only recently it has become a central player of it as 
the Organization for Economic Co-operation and Development
recognizes:
\begin{quote}
``The term ``knowledge-based economy'' results from a fuller recognition
of the role of knowledge and technology in economic growth. Knowledge,
as embodied in human beings (as ``human capital'') and in technology,
has always been central to economic development. But only over the
last few years has its relative importance been recognised, just as
that importance is growing. The OECD economies are more strongly
dependent on the production, distribution and use of knowledge than
ever before.''\footnote{
OECD.
The Knowledge-based economy.
OECD, Paris 1996.
}
\end{quote}


 From this statement   
and the premise  stated above by Jim Gray (``science today
is heavily based on data''),  the  conclusion follows:
data is the raw element of this new process of
production. A more allegorical version of this statement is:
``data has become the new oil''.\footnote{
Seems that Clive Humby was the first to coin this statement:
``Data is just like crude. It's valuable, but if unrefined it cannot really be used. It has to be changed into gas, plastic, chemicals, etc. to create a valuable entity that drives profitable activity; so must data be broken down, analyzed for it to have value.''
\url{http://ana.blogs.com/maestros/2006/11/data_is_the_new.html}
Since then, in Forbes, Fortune, Wired, etc. have appeared  articles with this
idea in the title.
  }

  As data is playing an essential  role in the economy, its production
process is being under pressure for efficiency.
Thus division of labor is affecting its cycle 
 --data capture; data curation; data analysis; data visualization--
 that traditionally was done by the
same person or team. (Tycho Brahe / Copernicus are an exception). 
The scientist and his collaborators designed the experiment or the
process of data collection (Von Humbolt, Darwin, Mendel, Pasteur, etc.).
 In particular, today there is an increasing tendency to separate the
uses and the production/collection of data.
In this way data is acquiring a certain degree of autonomy.

 
  Another relevant facet of scientific data is the old, but ongoing debate, about
  the epistemic status of observation versus experimentation. The first, rather direct
and implicitly without touching, without asking, the object.
The second is a product of direct manipulation of the object to
extract what is needed. Bogen gives a good example:
``To look at a berry on a vine and attend to its color and shape would be to observe it. To extract its juice and apply reagents to test for the presence of copper compounds would be to perform an experiment.''\footnote{
  J. Bogen. {\em op. cit.}
  } 
      The distinction, if there is one, is subtle.
One can state it in computational terms as the question: 
Static or Dynamic data? Bulk Data or API?~\footnote{
J. Tauberer. August 2014.
\url{https://opengovdata.io/2014/bulk-data-an-api/}
(API: Application Programmer Interface. For data, intuitively, an
interface oriented to be used not by a human, but as source
where applications can be pluged to automatically interact
with or retrieve data.)
}
s
The discussion is relevant not only for how to collect or
produce the data, but for how to store it and how to
deliver it to final users.
In fact we need both types of data.
Today we can ``expose'' live data, in the form of an
API, through cameras, from sensors, etc.  
This is becoming an increasingly relevant source of data. 
There is a growing interest
in technologies devised to process them lively, {\em i.e.}, as a stream
of data. Common examples are the value of currencies on the Web,
Weather channels, live news, etc.

Last, but not least, we should call the attention to
the blurring differences between ``scientific'' data and
``common'' data.
Data comes in many forms and  sources. 
One speaks of scientific data as that collected systematically 
in the framework of a scientific endeavor. Today, there are
huge data companies  outside what
we would consider ``scientific'' projects or institutions,
particularly at the social level (which are among
the most noisy and popular data). Tweets, identities and
behavior of users in social networks, social footprints of
any kind, personal images and videos, etc., are among
the most valuable data.
It is becoming everyday more difficult to trace a clear
divisory line between scientific and, say, non-scientific data.
At the end, all data is collected with some purpose in mind
(nobody would spend time, energy, resources to collect data
that  would not have some, although far off, goal.)

\section{The Social Character of Data}

  We have learned that data is everywhere; that data is relevant;
that it is valuable. Not surprisingly
international organizations, governments, communities are
devising ways to approach, and/or take advantage of, this new resource.


  As we saw, data is a resource that is essential to the development
of scientific knowledge, and as such, relevant to the understanding
of us as humans, to the development of our societies,
and to satisfy personal human needs. On the other hand, as
a ``new oil'', that is, as an economic good, it is under the
tension of economic categories.

%

 A naive approach would treat data in a similar way
as knowledge, a resource that looks at first sight as non-excludable
and non-sustractable, as Joseph Stiglitz, then Chief economist of the
World Bank, explained:

\begin{quote}
``A public good has two critical properties, non-rivalrous consumption--the consumption of one individual does
not detract from that of another--and non-excludability--it is difficult if not impossible to exclude an individual from
enjoying the good.
[...]
Knowledge is a global public good requiring public support at the global level.''.\footnote{
J. Stiglitz, Knowledge as a Global Public Good.
In: Global Public Goods: International Cooperation in
the 21st Century. 1998. 
}
\end{quote}

We could change ``knowledge'' by data and obtain a program for
data as a public good. It is, in fact, the program of several government
and international agencies. 
For example, the World Bank's focus is to make data
accesible to  particulars in order to
``allow policymakers and advocacy groups to make better-informed decisions and measure improvements more accurately.''\footnote{
World Bank Open Data Initiative.
World Bank, 4/30/2010. 
\url{data.worldbank.org}
}
   Along similar lines,
 OECD has a program for open access,
defined in its
{\em  Principles of Access to Research Data} as follows:
``Openness means access on equal terms for the international research
community at the lowest possible cost, preferably at no more than the
marginal cost of dissemination. Open access to research data from
public funding should be easy, timely, user-friendly and preferably
Internet-based.''~\footnote{
  OECD Principles and Guidelines for Access to Research
  Data from Public Funding. OECD, April 2007. p.15.
}
From these principles follow the transparency and interoperability policies
for governments.
 

  A good example of these initiatives in the scientific area
is  the U.S. {\em National Science Foundation}'s open data
policy, stating that
``agencies must adopt a presumption in favor of openness to the extent permitted by law and subject to privacy, confidentiality, security, or other valid restrictions.''\footnote{
Open Data at NSF. \url{https://www.nsf.gov/data/}
}
They define open data as follows:
\begin{quote}
``Open data are publicly available data structured in a way to be fully accessible and usable. This is important because data that is open, available, and accessible will help spur innovation and inform how agencies should evolve their programs to better meet the public's needs.''
\end{quote}
They state seven principles of consistency with open data, namely
to be public, accessible, described, reusable, complete, timely and 
managed post release.


   A different source for openness comes from
the pressure of diverse communities known as 
``open data'' movement.
Their notion of  open data is essentially taken 
from the ``open source'' and ``open access" communities. The ``translation'' of these notions to the world of data  bears 
the same  issues and challenges (no less, no more) than in those fields.
The {\em Open Data Handbook}~\footnote{
What is Open Data?   Open Data Handbook.
\url{http://opendatahandbook.org/guide/en/what-is-open-data/}
} 
   defines it as follows:
   ``Open data is data that can be freely used, re-used and redistributed by anyone --subject only, at most, to the requirement to attribute and sharealike.''
   As we can see, there is here a more ample conception than that of
 World Bank,  OECD and international organizations and goverments,
whose openness agendas are triggered mainly by economic concerns.
      
\medskip

\section{Final Remarks: Beyond Access}

Despite the advances these policies about access bring, relevant 
issues for data remain open.

  Most approaches used to address the notion of 
``open data'' implicitly associate it 
 to knowledge and/or information, whose  main threat is
effectively enclosure (in the form of patents and copyright).
A key assumption in this analysis is that the ``good'' 
under the threat of enclosure is something ready to be consumed.
Thus, the ultimate goal is access, that would allow people to 
consume that good.
This premise holds for simple data, 
as  spreadsheets,  transparency data, etc., but
does not hold for most data today, namely ``big'' data. 
 Access in this case is just a first step in the data cycle 
(collection; curation; analysis; visualization). 
The resources needed to store and cure data, to analyze,
and to finally visualize or use it, are tremendous. 
  The challenge is the scale. 

  Here the framework of commons comes to the rescue.
As Charlotte Hess and Elinor Ostrom state, 
\begin{quote}
``the essential questions for any commons analysis are 
inevitably about equity, efficiency and sustainability. 
Equity refers to issues of just or equal 
appropriation from, and contribution to, the maintenance of 
a resource.
Efficiency deals with optimal production, management and use of the
resource.
Sustainability looks at the oucomes over the long term.''\footnote{
  Ch. Hess, E. Ostrom (Ed.)
Understanding Knowledge as a Commons
From Theory to Practice. MIT Press 2006.
Introduction,  p. 6.
} 
\end{quote}

  Due to the enormous sizes of data,
to think about data as common implies including the whole cycle of
data as commons. Data is a resource shared (and produced) by 
groups of people. On its intangible face, is clearly non-excludable
and non-rivalrous: sharing it is almost effortless;
 consuming it does not substract the possibility
of others to do the same.
 The problem comes when we consider its material face.
Here  all the issues of a ``material'' commons surface, with 
its dilemmas of commodification or enclosure, pollution and
degradation, and nonsustainability.

Data came to our societies to stay.
And we already hear that data is the new oil. 
The allegory can be expanded to include the history of oil on earth:
It warns us of the possible conflicts that 
the appropriation of this new resource could bring.
It will depend on us, humans, to define what we want from this
new oil and how  we can use it to improve our lives and societies.

The discussion  we should open  today is how we 
would like to manage and govern this new good,
 including how it is generated, accessed,
stored, curated, processed and delivered. 
The commons approach offers  fresh insights to address these challenges.




\end{document}